\documentclass[aps,prd,twocolumn,superscriptaddress,showpacs]{revtex4-2}
\usepackage{graphicx}
\usepackage{amsmath}
\usepackage{amsfonts}
\usepackage{amssymb}
\usepackage{amssymb}
\usepackage{slashed}
\usepackage{pstricks,pst-coil}
\usepackage{dsfont}
\usepackage{bm}

\begin{document}
\title{Effective potential of scalar Lee-Wick pseudo-electrodynamics}
%
%


\author{M. J. Neves}\email{mariojr@ufrrj.br}

\affiliation{Departamento de F\'isica, Universidade Federal Rural do Rio de Janeiro, BR 465-07, 23890-971, Serop\'edica, Rio de Janeiro, Brazil}

\begin{abstract}
The study of effective potential for the scalar Lee-Wick pseudo-electrodynamics in one-loop is presented in this letter.
The planar and non-local Lee-Wick pseudo-electrodynamics is so coupled to a complex scalar field sector in $1+2$ dimensions, 
where we achieve the Lee-Wick pseudo-scalar electrodynamics. The effective action formalism is applied such that the quantum 
corrections are examined in one loop to the scalar effective potential as function of the classical field, of the Lee-Wick mass, 
and also of the coupling constants of this model. The instability of the effective potential is investigated due to Lee-Wick mass.

\end{abstract}
\maketitle

%
%



%

%
{\it Introduction} - The pseudo-electrodynamics (PED) has been investigated in the literature 
as the theory that mediates the interactions of electrons in material physics 
\cite{Marino93}. As examples, Dirac materials \cite{Castro}, topological planar 
materials \cite{Qi,Hasan,Chiu,Zhao,Qi2008}, superconductivity in layered materials 
\cite{Tesanovic,Zhang,Franz,Kivelson,Marino2018}, graphene \cite{Gorbar1,Herbut,Gusynin,Herbut2}  
and others systems show the good connection between condensed matter physics (CMP) and quantum field theories (QFTs) 
for the description of quantum phenomena. The PED is a non-local and planar theory in $1+2$ dimensions that 
preserves the basic properties of QFT as causality \cite{Amaral}, unitarity \cite{Marino2014}, and it is renormalized when 
coupled to fermions \cite{Alves2}. Several extensions of PED to include mass for the gauge field are discussed in the literature, 
as the pseudo-Proca ED \cite{Alves,Ozela,Ozela2,VanSergio}, that breaks the gauge symmetry, and the supersymmetric version of the PED \cite{PetrovPLB}. 
One of ways to keep the gauge invariance in the presence of massive degree of freedom is the study of a Lee-Wick electrodynamics reduced to $1+2$ dimensions \cite{MarioPRD2025}.
The Lee-Wick pseudo-ED (LWPED) is the version of the PED in the presence of a heavy mass in the gauge sector beyond the usual 
massless propagation \cite{lw69,lw70,podolsky42,podolsky44}. The coupling of the LWPED with fermions was showed that quantum corrections are finite 
in one loop \cite{MarioArxivAPR}. These results motivate us to investigation of quantum effects when the LWPED is coupled to scalar fields. 
The effective potential in the PED was obtained in the ref. \cite{Nascimento}.      
In this letter, we investigate the effective potential in one loop for the complex scalar theory coupled to the PLWED. We use the 
effective action functional approach to obtain the effective potential when the scalar complex field is expanded around 
a classical scalar field. The effective action and the correspondent potential in one loop is so obtained in terms 
of the operators in the propagation sector of the scalars and gauge fields. The LW mass plays a fundamental role in the instability 
of the effective potential. The letter is organized as follows : The first part is dedicated to the scalar Lee-Wick pseudo-ED and the 
functional formulation with the result of the effective potential. The second part describes the renormalization of the effective potential.  
For end, we end up with the conclusions. We start the paper with the natural units of $\hbar=c=1$, and in the perturbative approach for the effective 
potential, the $\hbar$-constant is reconsidered just in loop effects. The bar greek index $\bar{\mu}=(0,1,2)$ is used for the space-time $1+2$, 
with the metric $\eta^{\bar{\mu}\bar{\nu}}=(+1,-1,-1)$.

\vspace{0.3cm}

%

%
{\it The Lee Wick pseudo-scalar quantum ED} - A complex scalar field $\phi$ coupled to the LW pseudo-gauge field is set by the lagrangian density  
\begin{eqnarray} \label{Lsc}
\mathcal{L}_{sc} &=& 
|D_{\bar{\mu}}\phi|^{2}-V(\phi^{\ast},\phi)
+\frac{1}{M_{\phi}^2}|D_{\bar{\mu}}D^{\bar{\mu}}\phi|^2 \; ,
\end{eqnarray}
where the covariant derivative operator is $D_{\bar{\mu}}\phi=\partial_{\bar{\mu}}\phi+igA_{\bar{\mu}}\phi$, 
$\bar{\mu}=\{0,1,2\}$ means the bar space-time index in $1+2$ dimensions, $g$ is a dimensionless coupling constant, 
$A^{\bar{\mu}}$ is the gauge potential of the LW pseudo-ED, and $M_{\phi}$ is the Lee-Wick mass for the scalar field.
Since $M_{\phi}$ is a heavy mass, the limit $M_{\phi} \rightarrow \infty$ reduces (\ref{Lsc}) to the usual complex 
scalar field in $1+2$ dimensions. The scalar potential is defined by $V(\phi^{\ast},\phi)=\mu^2|\phi|^2+\lambda|\phi|^4$, in which 
$\mu$ and $\lambda$ are two real parameters, both with mass dimension in $1+2$ dimensions. The dynamics of the LW pseudo-gauge field is governed by \cite{MarioPRD2025} 
\begin{eqnarray} \label{LPLW}
\mathcal{L}_{PLW} = -\frac{1}{4} \, F_{\bar{\mu}\bar{\nu}}N(\bar{\Box})F^{\bar{\mu}\bar{\nu}} \; ,
\end{eqnarray}
where $F_{\bar{\mu}\bar{\nu}}=\partial_{\bar{\mu}}A_{\bar{\nu}}-\partial_{\bar{\nu}}A_{\bar{\mu}}=(-E_{x},-E_{y},B)$ is the EM tensor, 
the $N(\bar{\Box})$-operator is
\begin{eqnarray}\label{Nop}
N(\bar{\Box})=\frac{2(\bar{\Box}+M^2)}{(\bar{\Box}+M^2)\sqrt{\bar{\Box}}-\bar{\Box}\,\sqrt{\bar{\Box}+M^2}} \; ,
\end{eqnarray}
and the bar D'Alembertian operator is defined by $\bar{\Box}=\partial_{t}^2-\partial_{x}^2-\partial_{y}^2$.
Notice that in 1+2 dimensions, the magnetic field $(B)$ is a pseudoscalar.
In the limit $M \rightarrow \infty$, the $N$-operator is  $N(\bar{\Box})=\frac{2}{\sqrt{\bar{\Box}}}$, 
and the lagrangian density (\ref{LPLW}) reduces to the pseudo-ED presented in ref. \cite{Marino93}. With all these definitions, 
the addition of (\ref{Lsc}) with (\ref{LPLW}) shows a model $U(1)$-gauge invariant, that we call Lee-Wick pseudo-scalar electrodynamics.  
%

%
%

%
The scalar field is so shifted as
%
%
\begin{eqnarray}\label{phih}
\phi(\bar{x})=\frac{\phi_{c}+\phi_1(\bar{x})+i\phi_2(\bar{x})}{\sqrt{2}} \; ,
\end{eqnarray}
in which $\phi_{c}$ is a real constant field, $\phi_i(\bar{x}) \, (i=1,2)$ are real 
scalar fields as functions of the space-time coordinates $\bar{x}=x^{\bar{\mu}}=(t,x,y)$. 
These scalar fields have dimension of mass to the power 1/2 in $1+2$ dimensions. In QFT, the physical interpretation of $\phi_{c}$ 
is associated with the minimization of the classical potential $V(\phi^{\ast},\phi)$ at tree level in eq. (\ref{Lsc}), where the quantum corrections in 
$\hbar$ are not taken into account in this point. After the parametrization (\ref{phih}), the quadratic terms in the scalar sector are given by
\begin{eqnarray}\label{Leffh}
{\cal L}_{sc}^{(2)} &=& \frac{1}{2}\,\partial_{\bar{\mu}}\phi_1\left(1+\frac{\bar{\Box}}{M_{\phi}^2} \right)\partial^{\bar{\mu}}\phi_1
-\frac{1}{2} \, (\mu^2+3\lambda\,\phi_c^2) \, \phi_1^2
\nonumber \\
&&
\hspace{-0.6cm}
+\frac{1}{2}\,\partial_{\bar{\mu}}\phi_{2}\left(1+\frac{\bar{\Box}}{M_{\phi}^2} \right)\partial^{\bar{\mu}}\phi_2
-\frac{1}{2} \, (\mu^2+\lambda\,\phi_c^2) \, \phi_2^2
\; .
\end{eqnarray}
%
The $\phi_1$- and $\phi_2$-scalar fields have massive like terms of $\mu^2+3\lambda\,\phi_c^2$ and $\mu^2+\lambda\,\phi_c^2$, respectively, 
where the scalar Lee-Wick mass $M_{\phi}$ satisfies the condition $M_{\phi} > \sqrt{ \mu^2 + 3 \lambda \, \phi_c^2 }$. In fact, the scalar sector contains
two degrees of freedom with particles in which the mass $M_{\phi}$ is heavy in relation to masses of $\phi_{i}$. 
Thereby, the condition of $M_{\phi} \gg \sqrt{ \mu^2 + 3 \lambda \, \phi_c^2 }$ must be used throughout this letter, such that 
the scalar propagator in the momentum space $p^{\bar{\mu}}=(p_{0},p_{x},p_{y})$ is
\begin{equation}\label{propPhisimpl}
\Delta(\bar{p}) \simeq \frac{i}{\bar{p}^2-m_{\phi_i}^2}-\frac{i}{\bar{p}^2-M_\phi^2}\simeq\frac{- \, i \, M_{\phi}^2}{(\bar{p}^2-m_{\phi_i}^2)(\bar{p}^2-M_\phi^2)} \; ,                                                 
\end{equation}      
that in the ultraviolet regime has the behaviour of $\Delta(\bar{p}) \sim (\bar{p}^2)^{-2}$, with $\bar{p}^2=p_{\bar{\mu}}\,p^{\bar{\mu}}=p_{0}^2-p_{x}^2-p_{y}^2$. 
Therefore, we have a model where the radiative corrections are finite integrals already in one loop approximation for $1+2$ dimensions.
In the parametrization (\ref{phih}), the gauge sector has kinetic terms
\begin{equation} \label{PLWED}
\mathcal{L}^{(2)}_{PLW} = -\frac{1}{4} \, F_{\bar{\mu}\bar{\nu}}N(\bar{\Box})F^{\bar{\mu}\bar{\nu}} +\frac{1}{2}\, g^2\,\phi_c^2 \,   
A_{\bar{\mu}}A^{\bar{\mu}}
\; ,
\end{equation}
that breaks the $U(1)$-gauge symmetry, and is known as Proca-Lee-Wick pseudo-ED. Similarly to the scalar sector, the lagrangian density 
(\ref{PLWED}) contains a heavy spin-one degree of freedom $M$, and now a light spin-one of mass $g^2 \, \phi_c^2$ that must satisfy the condition $M \gg g^2 \, \phi_{c}^2$. 
After integrations by parts, the quadratic terms in the action of the scalar plus the gauge sector are written as 
\begin{eqnarray}\label{S2}
S^{(2)}[\phi_{i},A^{\bar{\mu}}]=\int d^{3}\bar{x}^{\prime}\,d^{3}\bar{x} \left[ -\frac{1}{2} \, \phi_{i}(\bar{x}^{\prime}) \, {\cal A}_{ij}(\bar{x}^{\prime},\bar{x};\phi_c) \, \phi_{j}(\bar{x}) \right]
\nonumber \\
+ \int d^{3}\bar{x}^{\prime}\,d^{3}\bar{x} \left[ \frac{1}{2} \, A_{\bar{\mu}}(\bar{x}^{\prime}) \, {\cal O}^{\bar{\mu}\bar{\nu}}(\bar{x}^{\prime},\bar{x};\phi_c) \, A_{\bar{\nu}}(\bar{x}) \right]
\, , \hspace{0.5cm}
\end{eqnarray}
where the operators ${\cal A}_{ij}(\bar{x}^{\prime},\bar{x};\phi_c)$ and ${\cal O}^{\bar{\mu}\bar{\nu}}(\bar{x}^{\prime},\bar{x};\phi_c)$ are defined 
in the coordinate space
\begin{subequations}
\begin{eqnarray}
{\cal A}_{ij}(\bar{x}^{\prime},\bar{x};\phi_c)&=&\left[\phantom{\frac{1}{2}}\!\!\!\!\!-\partial_{\bar{x}^{\prime}}^{\bar{\mu}}\partial_{\bar{x}\bar{\mu}}
+m_{\phi_{i}}^2(\phi_c)
\right.
\nonumber \\
&&
\left.
\hspace{-0.4cm}
+\frac{1}{M_{\phi}^2}\, (\partial_{\bar{x}^{\prime}}^{\bar{\mu}}\partial_{\bar{x}\bar{\mu}})^2 \right] \delta_{ij} \, \delta^{3}(\bar{x}^{\prime}-\bar{x}) \; ,
\\
{\cal O}^{\bar{\mu}\bar{\nu}}(\bar{x}^{\prime},\bar{x};\phi_c)&=& \left[\phantom{\frac{1}{2}}\!\!\!\!N(-\partial_{\bar{x}^{\prime}}^{\bar{\alpha}}\partial_{\bar{x}\bar{\alpha}})
(-\eta^{\bar{\mu}\bar{\nu}}\partial_{\bar{x}^{\prime}}^{\bar{\alpha}}\partial_{\bar{x}\bar{\alpha}}+\partial_{\bar{x}^{\prime}}^{\bar{\mu}}\partial_{\bar{x}}^{\bar{\nu}})
\right.
\nonumber \\
&&
\left.
\hspace{-0.4cm}
+g^2\phi^2_c\,\eta^{\bar{\mu}\bar{\nu}}
\right]\delta^{3}(\bar{x}^{\prime}-\bar{x}) \; ,
\end{eqnarray}
\end{subequations}
with $m_{\phi_{1}}^2(\phi_c)=\mu^2+3\lambda\phi_c^2$, and $m_{\phi_{2}}^2(\phi_c)=\mu^2+\lambda\phi_c^2$. In (\ref{S2}), we have redefined the fields as $\phi_{i} \rightarrow \sqrt{\hbar} \, \phi_{i}$ and $A^{\bar{\mu}} \rightarrow \sqrt{\hbar} \, A^{\bar{\mu}}$ where terms proportional to $\hbar^{1/2}$, $\hbar$ and at higher order of $\hbar$ are considered as quantum corrections beyond the one-loop approximation. Thus, only the quadratic terms of the action are relevant at one-loop.     
Using the functional formulation of QFT (see the ref. \cite{Peskin} for a review), the effective action at one loop for the $\phi_c$-constant field is $\Gamma[\phi_c]=\Gamma_{0}[\phi_c]+\hbar \, \Gamma_{1}[\phi_c]$, where $\Gamma_{0}[\phi_c]$ is the effective action at classical level, and $\Gamma_{1}[\phi_c]$ 
is the one loop correction given by
\begin{eqnarray}
\Gamma_{1}[\phi_c] &=& - \, \Omega \, V_{1}(\phi_c)=-\frac{i}{2} \left\{ \, \mbox{Tr} \ln\left[ \frac{ {\cal A}(\bar{x}^{\prime},\bar{x};\phi_c) }{{\cal A}(\bar{x}^{\prime},\bar{x};0)} \right] 
\right.
\nonumber \\
&&
\left.
+\mbox{Tr} \ln\left[ \frac{ {\cal O}^{\bar{\mu}\bar{\nu}}(\bar{x}^{\prime},\bar{x};\phi_c) }{{\cal O}^{\bar{\mu}\bar{\nu}}(\bar{x}^{\prime},\bar{x};0)} \right] \, \right\} \; ,
\end{eqnarray}
in which $\int d^{3}\bar{x}=\Omega$ is the volume of space-time $1+2$, 
$\mbox{Tr}$ denotes the trace operation in the coordinate space, and in the gauge sector also includes the trace over the index $(\bar{\mu},\bar{\nu})$. 
In the momentum space, the effective potential at one loop is set by the integrals
\begin{eqnarray}\label{V1int}
V_{1}(\phi_c) &=& -\frac{i}{2}\int\frac{d^3\bar{k}}{(2\pi)^{3}} \, \ln\left[\, 1-\frac{\lambda\phi_c^2}{\bar{k}^2(1-\bar{k}^2/M_{\phi}^2)-\mu^2} \, \right]
\nonumber \\
&&
\hspace{-0.5cm}
-\frac{i}{2}\int\frac{d^3\bar{k}}{(2\pi)^{3}} \, \ln\left[\, 1-\frac{3\lambda\phi_c^2}{\bar{k}^2(1-\bar{k}^2/M_{\phi}^2)-\mu^2} \, \right]
\nonumber \\
&&
\hspace{-0.5cm}
-\frac{i}{2}\int\frac{d^3\bar{k}}{(2\pi)^{3}} \, \ln\left[\, 1-\frac{3 g^2 \phi_c^2 
}{2\,N(-\bar{k}^2) \, \bar{k}^2} \, \right] \; ,
\end{eqnarray}
that are divergent in the ultraviolet regime. Thereby, the dimensional regularization $(D)$ is so introduced such that the previous integrations are modified by
\begin{eqnarray}\label{V1intD}
V_{1}(\phi_c,D) &=& -\frac{i}{2}\int\frac{d^{D}\bar{k}}{(2\pi)^{D}} \, \ln\left[\, 1-\frac{\lambda\phi_c^2}{\bar{k}^2\,(1-\bar{k}^2/M_{\phi}^2)-\mu^2} \, \right]
\nonumber \\
&&
\hspace{-0.5cm}
-\frac{i}{2}\int\frac{d^{D}\bar{k}}{(2\pi)^{D}} \, \ln\left[\, 1-\frac{3\lambda\phi_c^2}{\bar{k}^2\,(1-\bar{k}^2/M_{\phi}^2)-\mu^2} \, \right]
\nonumber \\
&&
\hspace{-0.5cm}
-\frac{i}{2}\int\frac{d^{D}\bar{k}}{(2\pi)^{D}} \, \ln\left[\, 1-\frac{3 g^2 (\Lambda)^{3-D} \phi_c^2
}{2\,N(-\bar{k}^2) \, \bar{k}^2} \, \right] \; ,
\end{eqnarray}
in which the physical result is recovered in the limit $D \rightarrow 3$. We have introduced the arbitrary energy scale $\Lambda$ to make the $g$-coupling constant dimensionless in $D$-dimensions. The integrations are evaluated in the Euclidian space, where $k_{0}=i\,k_{4}$, with $\bar{k}^2=-\bar{k}_{E}^{2}$ and $d^{D}\bar{k}=i\,d^{D}\bar{k}_{E}$. Using the known technical of the dimensional regularization \cite{Peskin}, the result of (\ref{V1intD}) is 
\begin{widetext}
\begin{eqnarray}\label{V1}
V_{1}(\phi_c,D) &=& M_{\phi}^{D} \, \frac{\Gamma(D/2)\,\Gamma(1-D/2)}{(8\pi)^{D/2} \, \Gamma(1+D/2)}\left[ \, 
\left(1+ \sqrt{1-4 \left(\frac{\mu^2+\lambda\phi_c^2}{M_{\phi}^2}\right)}\,\right)^{D/2}
\!\!\!\!\!+\left(1- \sqrt{1-4 \left(\frac{\mu^2+\lambda\phi_c^2}{M_{\phi}^2}\right) } \,\right)^{D/2}
\right.
\nonumber \\
&&
\left.
-\left(1+ \sqrt{1-\frac{4\mu^2}{M_{\phi}^2}}\,\right)^{D/2}
\!\!\!\!\!-\left(1- \sqrt{1-\frac{4\mu^2}{M_{\phi}^2}}\,\right)^{D/2}
\right]
\nonumber \\
&&
+M_{\phi}^{D} \, \frac{\Gamma(D/2)\,\Gamma(1-D/2)}{(8\pi)^{D/2} \, \Gamma(1+D/2)}\left[ \, 
\left(1+ \sqrt{1-4 \left(\frac{\mu^2+3\lambda\phi_c^2}{M_{\phi}^2}\right)}\,\right)^{D/2}
\!\!\!\!\!+\left(1- \sqrt{1-4 \left(\frac{\mu^2+3\lambda\phi_c^2}{M_{\phi}^2}\right)}\,\right)^{D/2}
\right.
\nonumber \\
&&
\left.
-\left(1+ \sqrt{1-\frac{4\mu^2}{M_{\phi}^2}}\,\right)^{D/2}
\!\!\!\!\!-\left(1- \sqrt{1-\frac{4\mu^2}{M_{\phi}^2}}\,\right)^{D/2}
\right]
\nonumber \\
&&
+\frac{M^{D}}{2^{D+1}(4\pi)^{D/2}\,\Gamma(1+D/2)}\left[ \, \frac{3g^2(\Lambda)^{3-D}\phi_{c}^2}{M} \, \right]^{D/3}\Gamma\left(1-\frac{D}{3}\right)\,\Gamma\left( \frac{D}{3} \right) \; ,
\end{eqnarray}
\end{widetext}
that is valid for the condition $0<\Re[D]<3$.

\pagebreak

Expanding the result (\ref{V1}) around $D=3-\epsilon$, for $\epsilon \rightarrow 0$, and using the approximations of 
$M_{\phi} \gg \mu \gg \lambda$, we obtain  
%
%

%
\begin{eqnarray}\label{V1result}
V_{1}(\phi_c,\epsilon) &=&
\frac{\lambda\,M_{\phi}}{\pi}\,\phi_c^2 
+ \frac{3g^2M^{2}\phi_c^2}{32\pi^2}\frac{1}{\epsilon}
\nonumber \\
&&
\hspace{-0.5cm}
+\frac{g^2M^2}{64\pi^2} \, \phi_c^2 \left[ \, 8-3\gamma-2\ln\left( \frac{3g^2M^2\phi_c^2}{8\pi^{3/2}\Lambda^{3}} \right) \, \right] 
\, , 
\hspace{0.8cm}
\end{eqnarray}
where $\gamma=0.577$ is the Euler-Mascheroni constant. The first integral in (\ref{V1intD}) is finite when $D=3$ and imposes 
the condition $M_{\phi}>2\sqrt{\mu^2+3\lambda\,\phi_c^2}$ for a real effective potential. The last term in the 
second line of (\ref{V1result}) diverges at $\epsilon=0$, that must be removed by a renormalization scheme.    
%


%
%

%
{\it The renormalized effective potential} - The emergence of a divergent term in the effective potential at one loop approximation points to need renormalization of the model.
Thereby, it is necessary to redefine the parameters of the model associated with this divergence. In particular, in the case of the effective potential, we introduce a counter-term that is quadratic in $\phi_c^2$, where the renormalized effective potential is given by   
\begin{eqnarray}\label{Veffrenor}
V_{eff}(\phi_c,\epsilon) &=& \frac{1}{2}\,\mu^2\phi_c^2+\frac{1}{4}\,\lambda\,\phi_c^{4}+\frac{1}{2}\,A\,\phi_c^2
\nonumber \\
%
&&
\hspace{-0.5cm}
+\frac{\lambda\,M_{\phi}}{\pi}\,\phi_c^2 + \frac{3g^2M^{2}\phi_c^2}{32\pi^2}\frac{1}{\epsilon}
\nonumber \\
&&
\hspace{-0.5cm}
+\frac{g^2M^2}{64\pi^2} \, \phi_c^2\left[ \, 8-3\gamma-2\ln\left( \frac{3g^2M^2\phi_c^2}{8\pi^{3/2}\Lambda^{3}} \right) \, \right] 
\; , \hspace{0.8cm}
\end{eqnarray}
in which the $A$-constant sets the counter-term, and now we have the finite effective potential of the model. 
The counter-term is determined by the renormalization condition 
\begin{eqnarray}\label{condd2V}
\left. \frac{d^2V_{eff}}{d\phi_c^2} \right|_{\phi_c=\sqrt{\Lambda}} \!\!=\mu^2 \; ,
\end{eqnarray}
where $\Lambda$ is the same arbitrary energy scale used to make $g$-coupling dimensionless. Using (\ref{condd2V}) in (\ref{Veffrenor}), we obtain
\begin{eqnarray}
A&=&-\frac{3g^2M^2}{16\pi^2}\frac{1}{\epsilon}-\frac{2\lambda M_{\phi}}{\pi}-3\lambda\,\Lambda
\nonumber \\
&&
-\frac{g^2M^2}{32\pi^2}\left[2-3\gamma-2\,\ln\left(\frac{3g^2M^2}{8\pi^{3/2}\Lambda^2} \right) \right]
\; .
\end{eqnarray}
The renormalized effective potential (in the limit $\epsilon \rightarrow 0$) is given by
%
\begin{eqnarray}\label{Veffrenorresult}
V_{eff}(\phi_c) &=& \frac{1}{2} \, \mu^2 \, \phi_c^2 +\frac{3g^2M^2}{32\pi^2} \, \phi_c^2 -\frac{4}{3} \, \lambda\,\Lambda \, \phi_c^2
\nonumber \\
&&
+\frac{\lambda}{4} \,\phi_c^{4}-\frac{g^2M^2}{32\pi^2} \, \phi_c^2 \, \ln\left( \frac{\phi_c^2}{\Lambda} \right)
\; .
\end{eqnarray}
%

\begin{figure}
\hspace{-0.5cm}
\includegraphics[width=1.0\linewidth]{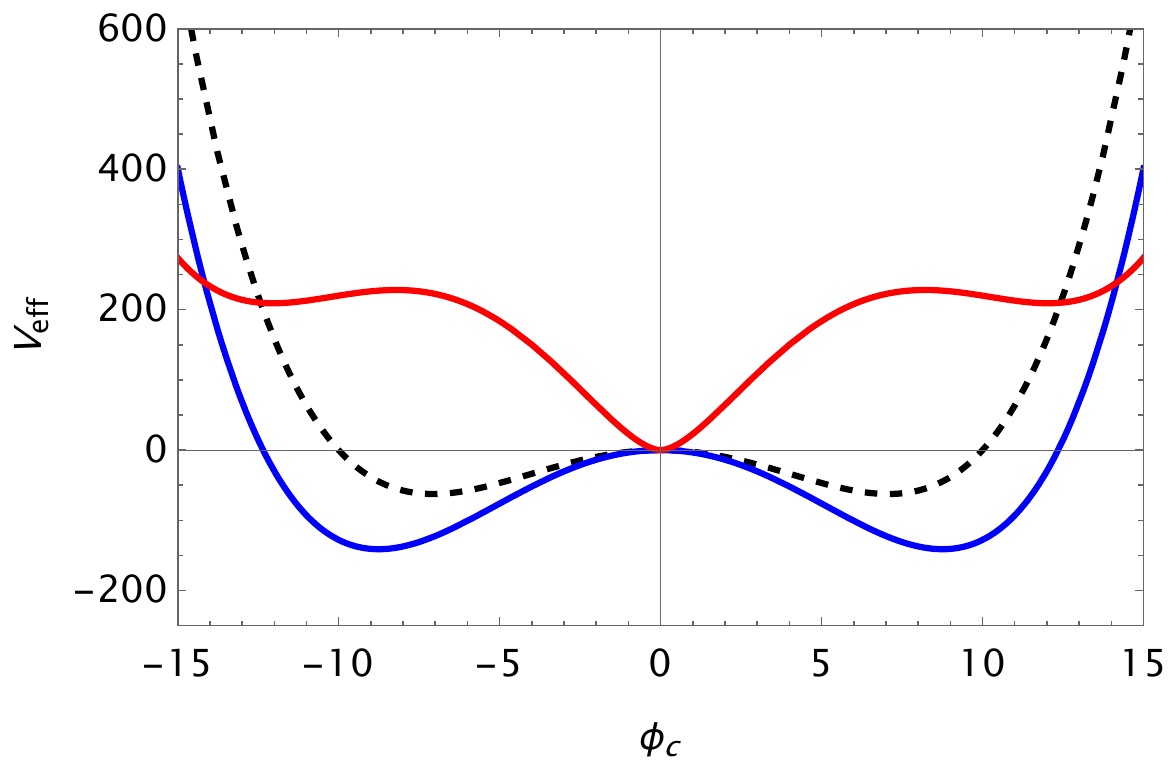}
\caption{The classical potential (dashed black line) and the renormalized effective potential (\ref{Veffrenorresult}) as functions 
of $\phi_c$ in $1+2$ dimensions. The blue line sets the case of renormalized effective potential with $M=50$, and the red line is illustrated 
for $M=400$ in mass units. We choose the values $\mu^2=-5$, $\lambda=0.1$, $\Lambda=10$ (all in energy units), and $g=0.1$.} \label{fig1}
\end{figure}
Notice that the scalar LW mass $M_{\phi}$ is removed by the renormalization, 
and the physical effective potential depends on gauge LW mass $M$. Since it was mentioned previously, 
the $\phi_c$-classical field is defined by the vacuum expectation value (VEV) scale $(v)$ 
as $\phi_c=\sqrt{v}$ through the minimization of the classical potential in the lagrangian (\ref{Lsc}). 
Thereby, the arbitrary scale $\Lambda$ can be calculated in terms of the VEV by the minimum condition 
in the effective potential
\begin{eqnarray}
\left. \frac{dV_{eff}}{d\phi_c} \right|_{\phi_c=\sqrt{v}} \!\!= 0 \; ,
\end{eqnarray}
that yields the result
\begin{eqnarray}
\Lambda=\frac{3g^2M^2}{128\pi^2\lambda}\, W\left[\frac{128\pi^2\lambda\,v}{g^2M^2}\right]\simeq v+\frac{3\lambda\,v^2}{g^2M^2} \; ,
\end{eqnarray}
where the $W$ is a product log function, and we have used the condition $M \gg v$. 
%
%


%
The result of effective potential (\ref{Veffrenorresult}) (blue line) in relation to the classical potential (dashed black line) is showed in the figure (\ref{fig1}).
The blue line means the renormalized effective potential for $M=50$, while the red line sets the case of LW mass $M=400$ (in energy units). In all these curves, we use the values of $\mu^2=-5$, $\lambda=0.1$, $\Lambda=10$, in energy units, and $g=0.1$ for the dimensionless coupling. For the case of $M=50$, the degenerated minimum of the effective potential function is at $\phi_{c}=\pm \, 8.75$ in squared root of energy unit, that sets the VEV scale of the model corrected by quantum effects,
with the vacuum energy of $-141.1$. The case of a heavier LW particle $M=400$ (red line) shows the renormalized effective potential with two points of 
instable equilibrium at $\phi_{c}=\pm \, 8.21$ for the maximum potential of $228.04$.   
%

%
%

%
%
  
%

%
%

%
%
%
%

%
%

%
%

{\it Conclusions} - We study the effective potential in one loop for the scalar Lee-Wick pseudo-electrodynamics (LWPED). The LWPED 
is the usual Lee-Wick ED reduced dimensionally from $1+3$ to $1+2$ dimensions with the classical sources constrained on the spatial plane $XY$.
The vantage of model is the inclusion of massive degree of freedom without the needed for a spontaneous symmetry breaking mechanism. Thereby, we have a non-local 
gauge symmetry in the presence of a heavy mass, that sets the LW particle.  Coupling this gauge field theory to a complex scalar field in the framework of LW, we apply the 
functional formulation to obtain the effective potential in one loop as function of the classical field, of gauge couplings, and of the LW masses. 
We obtain divergent contributions at one loop in $D=3$, where the unwanted term is removed by a subtraction scheme (renormalization). Therefore, we get a finite effective potential that is affected by the gauge LW mass. The figure \ref{fig1} shows the curves of the effective potential as function of the classical field for some values of the LW mass. For a LW mass $M=50$ (blue line), the effective potential  at one loop keeps the same shape from the classical potential (dashed line), with two stable points for $\mu^2 < 0$. For a heavier LW mass of $M=400$ (red curve) , the effective potential presents two unstable points with maximum of potential energy. The stable equilibrium point is at origin $\phi_c=0$. Thus, the LW has a range of values that can modify the shape of the effective potential. Furthermore, the LW mass works as a natural regulator for ultraviolet divergences in that the first quantum corrections are finite in $1+2$ dimensions that leads the scalar LWPED to be a super-renormalizable model.

\end{document}